# Nonlinear Dynamics and Chaos: Applications in Atmospheric Sciences

A.M.Selvam[1]

Deputy Director (Retired)
Indian Institute of Tropical Meteorology, Pune 411005, India
Email: amselvam@gmail.com
Web sites: http://www.geocities.ws/amselvam
http://amselvam.tripod.com/index.html
http://amselvam.webs.com

## Abstract

Atmospheric flows, an example of turbulent fluid flows, exhibit fractal fluctuations of all space-time scales ranging from turbulence scale of mm - sec to climate scales of thousands of kilometers – years and may be visualized as a nested continuum of weather cycles or periodicities, the smaller cycles existing as intrinsic fine structure of the larger cycles. The power spectra of fractal fluctuations exhibit inverse power law form signifying long - range correlations identified as self - organized criticality and are ubiquitous to dynamical systems in nature and is manifested as sensitive dependence on initial condition or 'deterministic chaos' in finite precision computer realizations of nonlinear mathematical models of real world dynamical systems such as atmospheric flows. Though the self-similar nature of atmospheric flows have been widely documented and discussed during the last three to four decades, the exact physical mechanism is not yet identified. There now exists an urgent need to develop and incorporate basic physical concepts of nonlinear dynamics and chaos into classical meteorological theory for more realistic simulation and prediction of weather and climate. A review of nonlinear dynamics and chaos in meteorology and atmospheric physics is summarized in this paper.

[1] Corresponding author address: (Res.) Dr.Mrs.A.M.Selvam, B1 Aradhana, 42/2A Shivajinagar, Pune 411005, India. Tel. (Res.) 09102025538194, email: amselvam@gmail.com

## 1. Introduction

The history of nonlinear dynamics and chaos begins with the brilliant original contribution of Henri Poincare[1] and specifically the derivation of the Poincare' map, and later with the advent of digital computers, of the findings of Ueda[2] and Feigenbaum[3] of the explosive and period-doubling routes to chaos, respectively. These findings, later complemented by the mathematics of global bifurcation theory[4] and analysis of observed chaotic data[5] set the stage for the well established theory which today is the new science motivating applications to atmospheric science.

Atmospheric flows, an example of turbulent fluid flows exhibits signatures of nonlinear dynamics and chaos, namely, self - similar fractal fluctuations of all space - time scales ranging from turbulence scale of mm - sec to climate scales of kms - years and may be visualized as a nested continuum of cycles or periodicities, the smaller cycles existing as fine scale structure of larger cycles. Fractal fluctuations are ubiquitous to dynamical systems in nature such as river flows, heart beat patterns, population dynamics, computer realizations of nonlinear mathematical models of dynamical systems, etc., and has been identified in all areas of science and human interest[6]. The power spectra of fractal fluctuations exhibit inverse power law form $f^{\alpha}$ where $f$ is the frequency and $\alpha$, the exponent. The frequency range over which $\alpha$ is a constant exhibits self-similarity or scale invariance, i.e., the fluctuation intensity (variance or amplitude squared) is a function of $\alpha$ alone and is independent of any other intrinsic property of the dynamical system such as its physical, chemical or any other characteristic. Scale invariance of space - time fluctuations of dynamical systems signifies long - range correlations or non - local connections and is identified as self - organized criticality[7]. The physics of the observed universal characteristics of fractal fluctuations indicate a common physical mechanism governing the space - time evolution of dynamical systems[8]. Therefore a general systems theory[9-11] where the model concepts are independent of the exact details, such as the chemical, physical, physiological, etc., properties of the dynamical systems will be applicable to dynamical systems in nature. Identification of the physics of self-organized criticality will enable quantification of the space - time growth pattern of dynamical systems such as atmospheric flows for predictability of future evolution of weather patterns. In this paper a review is given of the current status of application of recently developed concepts in the new multi - disciplinary science of nonlinear dynamics and chaos in meteorology and atmospheric physics. The paper is organized as follows. A brief history of the new science of nonlinear dynamics and chaos is given in Sec. 2. The identification of fractals and self-organized criticality in meteorology and atmospheric physics is summarized in Sec. 3. The current status of applications of nonlinear dynamics and chaos for weather prediction is given in Sec. 4. A list of mathematical and physical topics relating chaos theory applied to atmosphere sciences is given in Sec. 5. Discussions and conclusions are given in Sec. 6.

## 2. New Science of Nonlinear Dynamics and Chaos

### 2.1 Dynamical systems and fractal space-time fluctuations

Dynamical systems in nature, i.e., systems that change with time, such as fluid flows, heartbeat patterns, spread of infectious diseases, etc., exhibit nonlinear (unpredictable) fluctuations. Conventional mathematical and statistical theories deal only with linear systems and the exact quantification and description of nonlinear fluctuations was not possible till the identification in the 1970s by Mandelbrot[6,12], of the universal symmetry of *self-similarity*, i.e., *fractal geometry* underlying the seemingly irregular fluctuations in space and time[13,14]. Fractals, as the name implies, describe non-Euclidean objects generic to nature such as tree roots, tree branches, river basins, etc., which occupy only a part (fraction) of the traditional (Euclidean) 3 or 2 dimensions[15]. The study of self-similar space-time fluctuations generic to dynamical systems, now (since 1980s), belongs to the newly emerging multidisciplinary science of *nonlinear dynamics and* chaos[16] and deals with unified concepts for fundamental aspects intrinsic to the complex (nonlinear) and apparently random (chaotic) space-time structures found in nature. Scientific community at large will derive immense benefit in terms of new insights and development of powerful analytical techniques in this multidisciplinary approach to quantify basic similarities in form and function in disparate contexts ranging from the microscopic to the macroscopic scale.

The apparently random, noisy or irregular space-time signals (patterns) of a dynamical system, however, exhibit qualitative similarity in pattern geometry on all scales, a signature of space-time correlations. In general, the spatiotemporal evolution of dynamical systems trace a zigzag (jagged) pattern of alternating increase and decrease, associated with bifurcation or branching on all scales of space and time, generating wrinkled or folded surfaces in three dimensions. Representative examples for time series of some meteorological parameters are shown in Fig. 1. Physical, chemical, biological and other dynamical systems exhibit similar universal irregular space-time fluctuations. A fascinating aspect of patterns in nature is that many of them have a universal character[17].

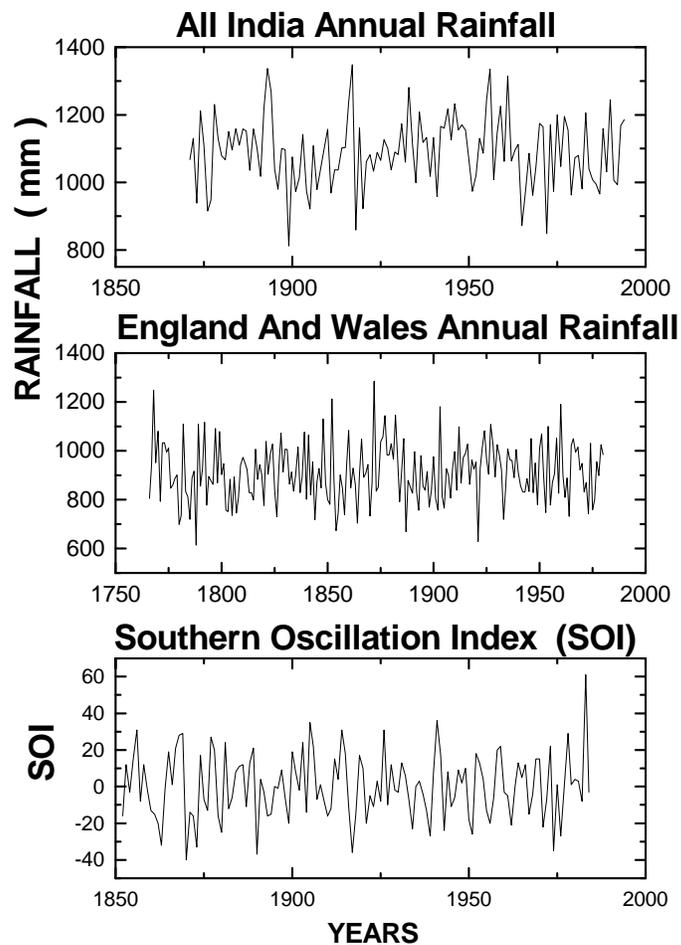

Fig. 1. Time series data of some of the meteorological parameters are shown as representative examples for irregular (zigzag) fluctuations (temporal) generic to dynamical systems in nature.

Irregular space-time fluctuations associated with basic bifurcation or branching geometry of wrinkles or folds on all scales is associated with the symmetry of self-similarity under scales transformation or just self – similarity[18]. A symmetry principle is simply a statement that something looks the same from certain different points of view. Such symmetries are often called principles of invariance[19]. The fundamental similarity or universality in the basic geometric structure, namely, irregularity, was identified as *fractal* in the late 1970s by Mandelbrot[6,12]. *Fractal* geometry is ubiquitous in nature, the fine structure on all scales being the optimum design for sustenance and growth of large-scale complex systems comprised of an integrated network of sub - units. The branching architecture of river tributaries, bronchial tree, tree branches, lightning discharge, etc., serve to collect/disperse fluids over a maximum surface area within a minimum volume. Fine-scale fluctuations help efficient mixing of fluids such as pollution dispersion in the atmosphere.

The basic similarity in the branching form underlying the individual leaf and the tree as a whole was identified more than three centuries ago in botany[20]. The importance of scaling concepts were recognized nearly a century ago in biology and botany where the dependence of a property $y$ on size $x$ is usually expressed by the allometric equation $y=ax^b$ where $a$ and $b$ are constants[11,21-23]. This type of scaling implies a hierarchy of substructures and was used by *D'Arcy Thompson* for scaling anatomical structures, for example, how proportions tend to vary as an animal grows in size[24,25]. *D'Arcy Thompson* (1963, first published in 1917)[21] in his book *On Growth and Form* has dealt extensively with similitude principle for biological modelling. Rapid advances have been made in recent years in the fields of biology and medicine in the application of scaling (*fractal*) concepts for description and quantification of physiological systems and their functions[23-28]. In meteorological theory, the concept of self - similar fluctuations was introduced in the description of turbulent flows by Richardson (1965, originally published in 1922)[29], Kolmogorov[30,31], Mandelbrot[32], Kadanoff[33] and others (see Monin and Yaglom[34] for a review).

## 2.2 Fractals in pure mathematics

Irregular (wrinkled) patterns are often described by functions that are continuous but not differentiable. Till the late 1800s pure mathematics dealt mostly with functions, which are differentiable everywhere such as the circle or ellipse. Pioneers in the study of functions which are continuous everywhere but without tangents are Karl Weierstrauss (1815-1897) who presented the Weierstrauss function in 1872, George Cantor (1845-1918) who provided the Cantor set in 1883 and Helge Van Koch (1906) who first constructed the snowflake curve[27]. A representative example, the Koch's curve is shown in Fig. 2.

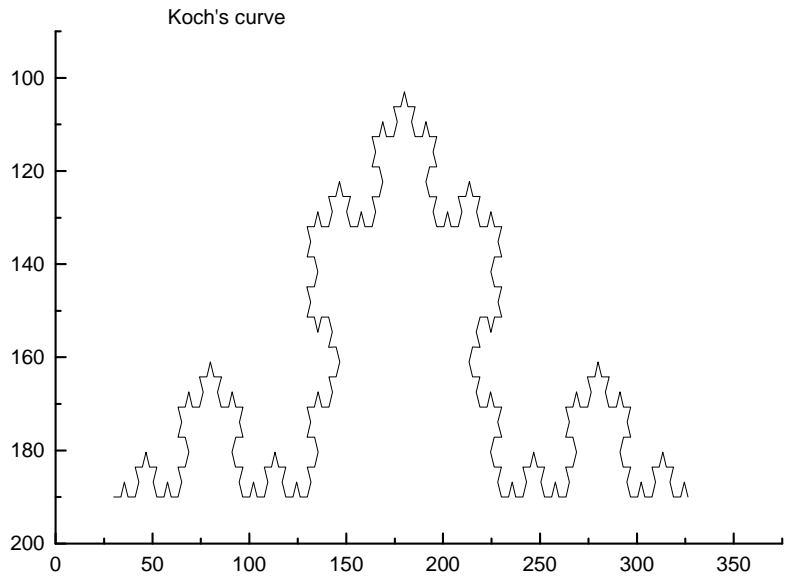

Fig. 2. The Koch's curve as a representative example for mathematical functions which are continuous everywhere but not differentiable anywhere, i.e., tangents cannot be drawn anywhere on the jagged boundary.

Jagged boundaries represented by these functions are more common in nature than the special case of curves with tangents, such as the circle. However, real world geometrical structures were not associated with these functions till a long time after their discovery. Continuous functions which are not differentiable anywhere represent an infinite number of zigzags between any two points. The length between any two points on the curve is infinity, yet the area bounded by the curve is finite. These "monster curves" which were outside the domain of pure mathematics were ignored as a field for study by many prominent mathematicians till the late 1800s.

The non-Euclidean geometry of the "monster curve" was quantified in terms of the similarity dimension by *Hausdorff* in 1919[35]. His idea was based on scaling, which means measuring the same object with different units of measurement. Any detail smaller than the unit of measurement is discarded. The jagged "monster curves" have fractional (non-integer) dimensions. The word *fractal* was coined by Mandelbrot[6] as a generic name for such objects as *Koch*'s snowflake, which possess fractional *Hausdorff* dimension. Besicovitch[36] was a second major figure who had developed the background for the concept of fractional dimension. Some of the earlier studies on applications of scaling concepts are given in the following. The question of scaling and the paradigm of *fractals*, i.e., when can a part have the same properties as the whole was addressed in the 1920s and 1930s by Levy[37] who was concerned with the question of when a sum of identically distributed random variables has the same probability distribution as any one of the terms in the sum[38]. The length of a *fractal* object, e.g., the coastline increases with decrease in the length of yardstick used for the measurement. Richardson[39] came close to the concept of *fractals* when he noted that the estimated length of an irregular coastline or boundary $B(l)$, where $l$ is the measuring unit is given by $B(l)=B_o l^{1-d}$ where $B_o$ is a constant with dimension of length and $d$ is the *fractal* dimension greater than 1 but less than 2 for the jagged coastline[24,25]. One of the oldest scaling laws in geophysics is the *Omori* law[40]. This law describes the temporal distribution of the number of after-shocks, which occur after a larger earthquake (i.e., main-shock) by a scaling relationship. The other basic empirical seismological law, the *Gutenberg-Richter* law[41] is also a scaling relationship, and relates intensity to its probability of occurrence[42]. The power-law is a distinctive experimental signature seen in a wide variety of complex systems. In economics, it goes by the name of 'fat tails', in physics it is referred to as 'critical fluctuations', in computer science and biology it is 'the edge of chaos', and in demographics it is called Zipf's law[43].

The *fractal* dimension $D$ in general for length scale $R$ may be given as:

$$D = \frac{\mathrm{d}\ln M}{\mathrm{d}\ln R}$$

where $M$ is the mass contained within a distance $R$ from a point in the extended object. A constant value for $D$ implies uniform stretching on logarithmic scale, resulting in large-scale structures which preserve their original geometrical shape. Objects in nature are in general *multi-fractals*, i.e. the *fractal* dimension $D$ varies with the length scale $R$. The *multi-fractal* nature of fluid turbulence and scaling concepts have been discussed by Sreenivasan[44].

The dimension of a naturally occurring *fractal* is a quantitative measure of a qualitative property of a structure that is self-similar over some regions of space or intervals of time. The powerful concept of fractal dimension introduced by Mandelbrot[6] has helped identify the universal symmetry of self-similarity underlying the seemingly irregular complex structures found in nature[13].

In summary, it is now accepted that dynamical systems in nature exhibit irregular space-time fluctuations. The geometrical structure of such *fractal* fluctuations is non-Euclidean and has fractional (non-integer) dimension. In this context a brief description of the concepts quantifying geometrical structures in traditional mathematics is given in the following. Classical Euclidean geometry deals only with regular objects such as *point*, *line*, *square* and *cube* in terms of integer dimensions *zero*, *one*, *two* and *three* respectively. The real world geometrical space is three-dimensional restricted to three mutually perpendicular directions (the Cartesian coordinates x, y and z). The concept of time is included separately as the fourth dimension in the description of evolution processes of three-dimensional real world systems. However, mathematical models of real world dynamical systems can have more than three dimensions, the dimensions in this case correspond to the number of degrees of freedom of the system under consideration. The degrees of freedom refer to the independent variables used in the mathematical model, e.g., the flight path of an aeroplane is given by six independent variables, or degrees of freedom, namely, the speed and momentum in the three mutually perpendicular directions at any instant.

## 2.3 Fractal fluctuations and statistical analysis

Most quantitative research involves the use of statistical methods presuming *independence* among data points and Gaussian 'normal' distributions[45]. The Gaussian distribution is reliably characterized by its stable mean and finite variance[46]. Normal distributions place a trivial amount of probability far from the mean and hence the mean is representative of most observations. Even the largest deviations, which are exceptionally rare, are still only about a factor of two from the mean in either direction and are well characterized by quoting a simple standard deviation[47]. However, apparently rare real life catastrophic events such as major earth quakes, stock market crashes, heavy rainfall events, etc., occur more frequently than indicated by the normal curve, i.e., they exhibit a probability distribution with a *fat tail*. Fat tails indicate a power law pattern and interdependence. The "tails" of a power-law curve — the regions to either side that correspond to large fluctuations

— fall off very slowly in comparison with those of the bell curve[48]. The normal distribution is therefore an inadequate model for extreme departures from the mean.

Fractals are the latest development in statistics. The space-time fluctuation pattern in dynamical systems was shown to have a self-similar or fractal structure. The larger scale fluctuation consists of smaller scale fluctuations identical in shape to the larger scale. An appreciation of the properties of fractals is changing the most basic ways we analyze and interpret data from experiments and is leading to new insights into understanding physical, chemical, biological, psychological, and social systems. Fractal systems extend over many scales and so cannot be characterized by a single characteristic average number[49]. Further, the self-similar fluctuations imply long-range space-time correlations or interdependence. Therefore, the Gaussian distribution will not be applicable for description of fractal data sets. However, the bell curve still continues to be used for approximate quantitative characterization of data which are now identified as fractal space-time fluctuations.

## 2.4 Golden mean and self-similar, *fractal* geometrical structures in nature

Animate and inanimate structures in nature exhibit self-similarity in geometrical shape[11,50-52], i.e., parts resemble the whole object in shape. The most fundamental self-similar structure is the forking (bifurcating) structure[11] of tree branches, tree roots, river tributaries, branched lightning, etc. The complex branching architecture is a self-similar *fractal* since branching occurs on all scales (sizes) and forms the geometrical shape of the whole object. Self-similar structures incorporate in their geometrical design the *noble numbers*, i.e., numbers, which are functions of the *golden mean* $\tau$ and are characterized by five-fold symmetry of the pentagon and dodecahedron. For example, the ratio of the length of the diagonal to the side in a regular pentagon is equal to the *golden mean* $\tau$ equal to $(1+\sqrt{5})/2 \cong 1.618$. The *golden mean* is the most irrational number and is associated with the *Fibonacci* mathematical sequence 1, 1, 2, 3, 5, 8,.. where, each term is the sum of the two previous terms and the ratio of each term to the previous term approaches the *golden mean* $\tau$. The *golden mean* $\tau$ is the most irrational number in the sense that rational approximations converge very slowly to $\tau$ as compared to other irrational numbers. Irrational numbers are numbers such as $\sqrt{2}$, which has an infinite number of non-periodic decimals. Rational approximations such as $p/q$ where $p$ and $q$ are integers are used to represent irrational numbers. The *golden mean* had a special significance in ancient cultures. The significance of the *golden mean* throughout recorded history in science, culture and religion has been discussed[53,54]. Self-similar spiral structures such as on the shell of the very old mollusk called *Nautilus* pompilius[11] incorporate the *golden mean* in their radial growth. *Thompson* described that the *nautilus* followed a pattern originally described by *Rene Descartes* in 1683 as the equiangular spiral and subsequently by *Jacob Bernoulli* as the logarithmic spiral[24]. The commonly found shapes in nature are the helix and the dodecahedron[55,56], which are signatures of self-similarity underlying *Fibonacci* numbers. The association of *noble numbers* with growth of self-similar patterns has been established quantitatively in plant *phyllotaxis* in *botany*. A summary of documented evidence collected over a period of more than 150 years is given below and will help understand the association between *noble numbers* and self-similar patterns in the plant kingdom. *Phyllotaxis* is the study of the arrangement of all plant elements, which originate as primordia on the shoot apex. The botanical elements, which constitute plants, are branches, leaves, petals, stamens, sepals, florets, etc. These plant elements begin their existence as primordia in the neighborhood of the undifferentiated shoot apex (extremity). Extensive observations in botany show that in more than 90% of plants studied worldwide[11,57] primordia emerge as protuberances at locations such that the angle subtended at the apical centre by two successive primordia is equal to the *golden angle* $\varphi = 2\pi \ (1-1/\tau)$ corresponding to approximately 137.5 degrees. Theoretical studies show that outside the set of *noble numbers* the structures are not self-similar. The surprisingly precise geometrical placement of plant primordia results in the observed *phyllotactic* patterns, namely, the familiar spiral patterns found in the arrangement of leaves on a stem, in florets of composite flowers, the pattern of scales on pineapple and pine cone, etc. Further, such self - similar patterns ensure identical geometrical design (shape) for all sizes of a single species such as daisy flowers of all sizes. The *phyllotactic* patterns, while pleasing to the eye, also incorporate maximum packing efficiency for fruits and seeds.

### 2.5 *Fibonacci* sequence and self-similar structures

The *Fibonacci* mathematical series was discovered in 1209 by Leonardo of Pisa, known as Fibonacci[58] while computing the total number of adult rabbits in successive months starting with a single adult rabbit pair and assuming that each adult rabbit pair produces one pair of offspring each month and that baby rabbit pairs became adults in one month's time. The growth of rabbit population is shown as a branching network in Fig. 3.

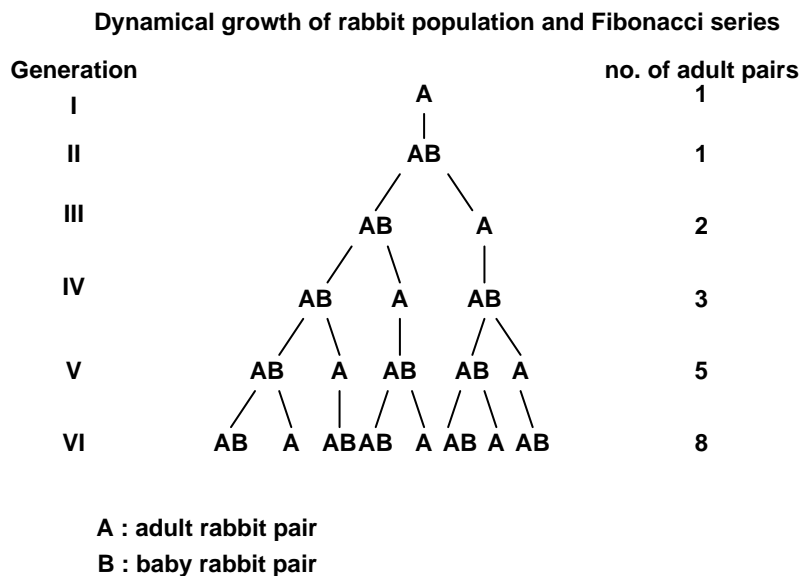

**Dynamical growth of rabbit population and Fibonacci series**

A : adult rabbit pair
B : baby rabbit pair

Fig. 3. Generation of *Fibonacci* numbers as cumulative sum of a sequence of ordered bifurcations (branchings).

The total number of adult rabbit pairs in successive months follows the *Fibonacci* mathematical series. The growth of adult rabbit population as shown in Fig. 3 represents a hierarchical ramified network or a self-similar *fractal* network. Ramified branching network systems in nature can be similarly shown to generate the *Fibonacci* mathematical number series. For example, the branching network of updrafts and downdrafts in vortex roll circulations in atmospheric flows (Fig. 4.) is shown to be represented by a hierarchy of branches with multiple sub-branches.

**Bifurcating network of Up and Downdrafts and Fibonacci numbers**

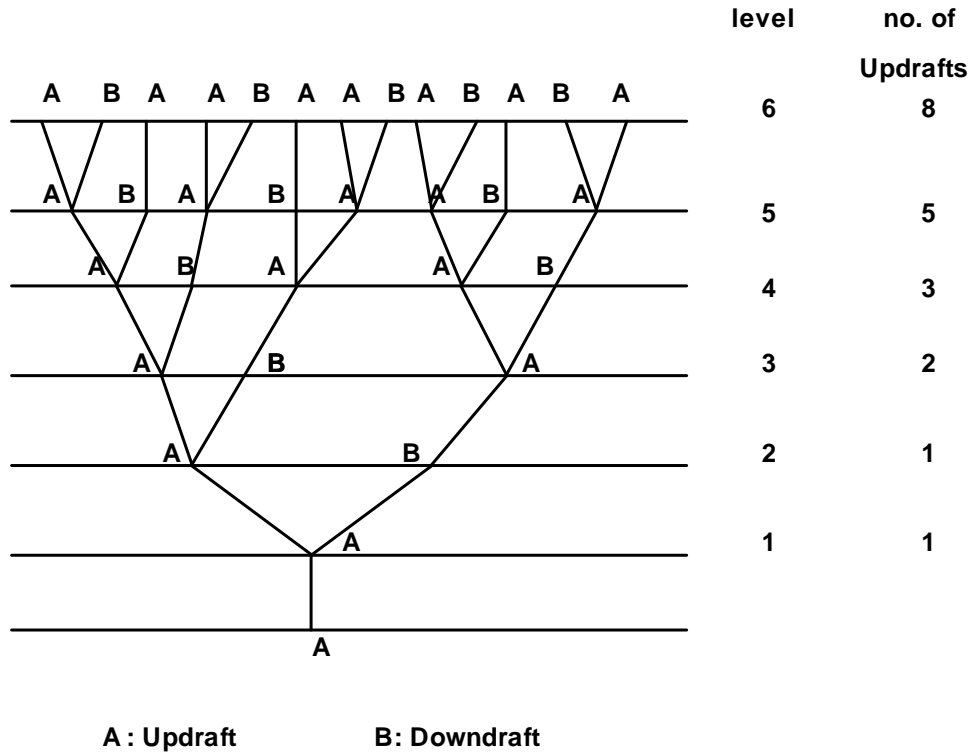

A : Updraft            B: Downdraft

Fig. 4. Bifurcating network of updrafts and downdrafts and *Fibonacci* numbers

In Fig. 4, A represents an updraft. At the first level, forking structure AB is generated with formation of sub-branch (downdraft) B. At the second level, A again generates the forking structure AB, while the sub-branch (downdraft) B of level 1 now generates the updraft A. Updrafts alone produce forking structure with formation of sub-branch (downdraft) B which then gives rise to updraft A at the next level. Continuing such a system of bifurcation results in the generation of *Fibonacci* numbers sequence for the total number of updrafts (A) at each level.

In summary, the integrated sum of smaller scale networks contribute to form large scale networks. Branching networks may therefore be considered as a hierarchy of self-similar networks or *fractals*. *Fractal* architecture to the spatial pattern is therefore a signature of cumulative integration (summation) process inherent to dynamical growth processes of the system. For example, the *fractal* network of a river drainage basin serves to collect water from the smallest of tributaries (branches) and integrate it into the main river flow.

### 2.6 Fivefold and spiral symmetry associated with Fibonacci sequence

The ratio of adjacent elements of the *Fibonacci* sequence approaches the irrational number $\tau$ = $(1+\sqrt{5})/2$ in the limit. The number, $\tau$, is the solution to the algebraic equation

$$1 + x = x^2$$

As a result $\tau$ has the property

$$1 + \tau = \tau^2$$

Therefore, the double geometric sequence...... $\frac{1}{\tau^3}, \frac{1}{\tau^2}, \frac{1}{\tau}$, 1, $\tau, \tau^2, \tau^3$,....... is the *Fibonacci* sequence since it has the property that each term is equal to the sum of the earlier two terms and also the ratio of each term to the earlier term is equal to the *golden mean* $\tau$. It is the only geometric series, which is also a *Fibonacci* sequence[58]. The *Fibonacci* numbers can be represented geometrically in polar coordinates in two dimensions by the equiangular spiral $R_OR_1R_2R_3R_4R_5$... drawn with origin O, with lengths of successive radii $OR_O$, $OR_1$, $OR_2$, ......... and corresponding spiral segments $R_OR_1$, $R_1R_2$, $R_2R_3$, ...... following *Fibonacci* mathematical sequence (Fig. 5).

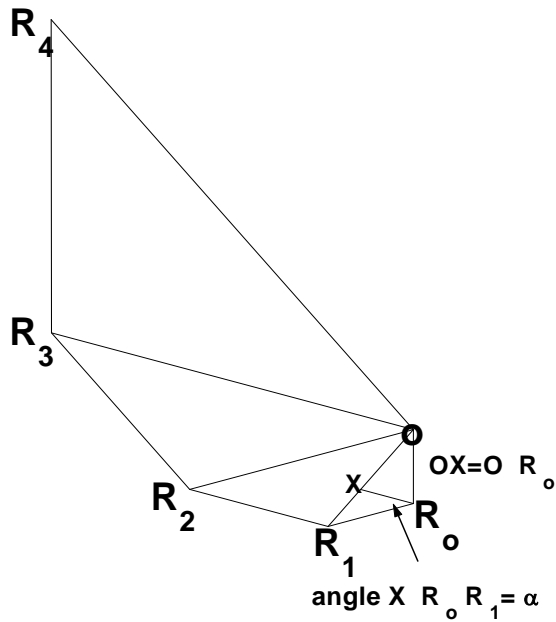

Fig. 5. Geometrical representation of *Fibonacci* series in polar coordinates by the *Fibonacci* equiangular spiral

$R_OR_1$ may be considered to be the tangent at $R_O$ to the circle with centre O and radius $OR_O$. The angle $R_1R_OX$ equal to $\alpha$ which the tangent makes with the arc $XR_O$ is the crossing angle of the spiral $R_OR_1R_2$...

The initial radius $OR_O$ equal to $R$ grows to $OR_1$ equal to $R + dR$ after an angular turning $R_OOR_1$ equal to $d\theta$. The incremental growth $dR$ is equal to the length $R_1X$ in Fig. 5.

Therefore $\qquad \tan\alpha = \dfrac{R_1X}{XR_O}$

$XR_O = Rd\theta$ for the arc $XR_O$ of circle with center O and radius $OR_O$ equal to $R$.

$\tan\alpha = \alpha$ in the limit for small values of $\alpha$

Therefore

$$\frac{\mathrm{d}R}{R\mathrm{d}\theta} = \alpha$$

or $\quad \dfrac{\mathrm{d}R}{R} = \alpha\,\mathrm{d}\theta$

$$\mathrm{d}\ln R = \alpha\,\mathrm{d}\theta$$

Integrating for growth of radius from $r$ to $R$ associated with angular turning from 0 to $\theta$,

$$\ln\frac{R}{r} = \alpha\theta$$

or

$$R = r\,\mathrm{e}^{\alpha\theta}$$

Geometrical consideration for generation of the *Fibonacci* spiral in three dimensions specify a constant angular turning $\mathrm{d}\theta$ equal to $\dfrac{1}{\tau}$ between successive radii and therefore a constant crossing angle, also equal to $\dfrac{1}{\tau}$. The *Fibonacci* equiangular spiral is then given by the relation

$$R = r\,\mathrm{e}^{\theta/\tau^2}$$

The angle subtended at the center between two successive radii is therefore equal to the golden angle $\quad \dfrac{2\pi}{\tau^2} \quad or \quad 2\pi(1 - \dfrac{1}{\tau}) \quad$ since $\quad \dfrac{1}{\tau^2} = 1 - \dfrac{1}{\tau}$

The *Fibonacci* equiangular spiral as shown in Fig. 6 has intrinsic internal structure of the quasiperiodic *Penrose tiling pattern* and associated long-range spatial and temporal correlations.

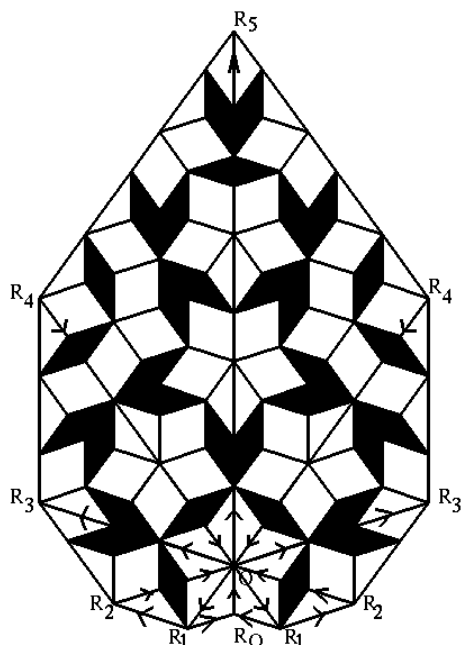

Fig. 6. Quasicrystalline structure of the quasiperiodic
*Penrose tiling pattern* and *Fibonacci* sequence.

The *Fibonacci* spiral is traced with mathematical precision in nature in the dynamical growth processes of plants as seen in the geometrical placement on the shoot, of primordia, which later develop into the various plant parts. In a majority (92%) of plants studied world - wide, successive primordia always subtend angle equal to the *golden angle* at the apical center[11].

Primordia placement in space and time may therefore be resolved into the precise geometrical pattern of the quasiperiodic *Penrose tiling* pattern.

## 2.7 Quasicrystalline structure: The quasiperiodic penrose tiling pattern

The regular arrangement of plant parts resembles the newly identified (since 1984) *quasicrystalline* order in condensed matter physics[59-61]. Traditional (last 100 years) crystallography has defined a crystalline structure as an arrangement of atoms that is periodic in three dimensions. Crystals have lattice structure with identical arrangement of atoms[62,63] with space filling cubes or hexagonal prisms. Five-fold symmetry was prohibited in classical crystallography. In 1984, an alloy of aluminum and magnesium was discovered which exhibited the symmetry of an icosahedron with five-fold axis. At the same time Paul Steinhardt of the University of Pennsylvania and his student Dov Levine[62] had quite independently identified similar geometrical structure, now called quasicrystals[64,65]. These developments were based on the important work on the mathematics of tilings done by *Roger Penrose* and others beginning in the 1970s. Penrose[66,67] discovered a nonperiodic tiling of the plane, using two types of tiles, which is a quasiperiodic crystal with pentagonal symmetry[68]. It is generally accepted that a quasicrystal can be understood as a systematic (but not periodic) filling of space by unit cells of more than one kind. Such extended structures in space can be orderly and systematic without being periodic. *Penrose tiling* patterns (Fig. 6) are two-dimensional quasicrystals.

The geometric pattern is self-similar and exhibits long-range correlations and is quasiperiodic. Selvam[69] has shown that turbulent fluid flows can be resolved into the quasiperiodic *Penrose tiling* pattern with *fractal* self-similar geometry to spatial pattern and long-range temporal correlations for temporal fluctuations. Self-organized criticality is exhibited as the *Penrose tiling* pattern for self-similar spatial geometry, which then incorporates temporal correlations for dynamical processes.

## 2.8 *Fractal* time signals, and power-laws

There are numerous power-law relations in science that have the self-similarity property. For example, the inverse square - law force, which is fundamental in gravitation and in electricity and magnetism, has no intrinsic scale. It has the same form at all scales under a linear scaling transformation[19,27]. The concept of *fractals* may be used for modelling certain aspects of dynamics, i.e., temporal evolution of spatially extended dynamical systems.

Spatially extended dynamical systems in nature exhibit *fractal* geometry to the spatial pattern and support dynamical processes on all time scales, for example, the *fractal* geometry to the global cloud cover pattern is associated with fluctuations of meteorological parameters on all time scales from seconds to years. The temporal fluctuations exhibit structure over multiple orders of temporal magnitude in the same way that *fractal* forms exhibit details over several orders of spatial magnitude. The power spectra of such broadband fluctuations exhibit inverse power-law of form $1/f^\alpha$ where $f$ is the frequency and $\alpha$ the exponent. In general, $\alpha$ decreases with $f$ and approaches 1 for low frequencies. Self-similar variations on different time scales will produce a frequency spectrum having an inverse $(1/f)$ power-law distribution or $1/f$ - like distribution and imply long-range temporal correlations signifying persistence or "memory". The frequency range over which $\alpha$ is constant therefore exhibits self-similarity or scale invariance in temporal fluctuations, i.e., the fluctuations are fractals in time. The intensity or variance of longer and shorter period fluctuations are mutually related by a scale factor alone independent of the nature of dynamical processes. The fluctuations exhibit long-range temporal correlations. Also, temporal fluctuations exhibit multifractal structure since $\alpha$ varies for different ranges of frequency $f$.

The phenomenon of $1/f$ - noise spectrum first introduced by Van Der Ziel in 1950[70,71] is ubiquitous to dynamical systems in nature and has a long history of more than 40 years of observational documentation in all fields of science and other areas[7,72-74]. The multidisciplinary nature of investigations will help gain new insights and develop mathematical and statistical techniques and analytical tools for understanding and quantifying the physics of the observed long-range correlations in dynamical systems in nature. The physics of dynamical systems therefore comes under the broad category of general systems theory. The subunits of the system function as a unified whole two-way communication and control network with global (system level) control/response to local functions/stimuli, thereby possessing the criteria for a robust system[75-77]. Kitano[76] makes the point that robustness is a property of an entire system; it may be that no individual component or process within a system would be robust, but the system - wide architecture still provides robust behavior. This presents a challenge for analysis, since elucidating such behaviors can be counterintuitive and computationally demanding.

Power - law behavior has been documented in the functioning of physiological systems[78,79]. Hurst[80] and Hurst et al[81] had shown for river flows[82] that for a wide variety of data sets the degree of "memory" over time spans of up to a millennium could be characterized by a power - law relationship[83]. Long - range spatial correlations have been identified at the level of the DNA[84,85]. Long - range correlations over time and space have also been investigated by Mandelbrot and Wallis[86] for geophysical records and more recently by Tang and Bak[87], and Bak et al[7,88], for $1/f$ noise in dynamical systems. Andriani and McKelvey[45] have given exhaustive references to earliest known work on power law relationships. A power law world is dominated by extreme events ignored in a Gaussian-world. In fact, the fat tails of power law distributions make large extreme events orders – of - magnitude more likely. Theories explaining power laws are also scale - free. This is to say, the same explanation (theory) applies at all levels of analysis[45].

The *1/f* power law would seem to be natural and white noise (flat distribution) would be the subject of involved investigation[71]. Recent studies have identified turbulent cascades in foreign exchange markets[89] and power - laws governing epidemics have been reported[90]. Universality gives a new understanding of how apparently very different things can act in the same way[91].

A major feature of this correlation is that the amplitude of short-term and long-term fluctuations are related to each other by the scale factor alone independent of details of growth mechanisms from smaller to larger scale. The macroscopic pattern, comprised of a multitude of sub-units, functions as a unified whole independent of details of dynamical processes governing its individual sub-units[92]. Such a concept that physical systems, which consist of a large number of interacting sub-units, obey universal laws that are independent of the microscopic details is now acknowledged as a breakthrough in statistical physics. The variability of individual elements in a system act cooperatively to establish regularity and stability in the system as a whole[71]. Scale invariance implies, knowledge of the properties of a model system at short times or short length scales can be used to predict the behaviour of a real system at large times and large length scales[93].

The *fractal* dimension *D* of a temporal *fractal* can be computed using recently developed algorithms. Since time series of a single variable such as temperature in atmospheric flows may reflect the cumulative effect of the multitude of factors governing flow dynamics, the *fractal* dimension may indicate the number of parameters controlling the evolution dynamics. However, a knowledge of *D* alone does not help identification of the parameters or their exact role in the dynamical growth processes. Also, limitations in data length and computational algorithms preclude exact determination of $D^{94}$. It has not been possible to formulate governing equations based on a knowledge of *D* for prediction purposes.

## 2.9 Self-organized criticality: space-time *fractal*s

Till very recently (1987), fractal geometry to the spatial pattern and fractal fluctuations in time of dynamical processes of the same extended dynamical system were treated as two disparate multidisciplinary fields of research[95]. The long-range spatiotemporal correlations underlying spatial and temporal power-law behavior of dynamical systems was identified as a unified manifestation of self-organized criticality (SOC) in 1987[7,14,88,95].

The unifying concept of self-organized criticality underlying fractals, self-similar scaling, broadband frequency spectra and inverse power-law distribution offer new and powerful means of describing certain basic aspects of spatial form and dynamical processes in a dynamical system. The systems in which self-organized criticality is observed range from the physical to the biological to the social. The physical mechanism underlying the observed self-organized criticality is not yet identified. However, the long-range spatial and temporal correlations underlying dynamical evolution implies predictability in space and time of the pattern of evolution of the dynamical system, for example, atmospheric flows.

The relation between spatial and temporal power - law behaviour was recognized much earlier in condensed matter physics where long - range spatiotemporal correlations appear spontaneously at the critical point for continuous phase transitions. The amplitudes of large and small-scale fluctuation are obtained from the same mathematical function using appropriate scale factor, i.e. ratio of the scale lengths. This property of self - similarity is often called a renormalization group relation in physics[24,96,97] in the area of continuous phase transitions at critical points[19,98]. When a system is poised at a critical point between two macroscopic phases, e.g., vapour to liquid, it exhibits dynamical structures on all available spatial scales, even though the underlying microscopic interactions tend to have a characteristic length scale[98]. But, in order to arrive at the critical point, one has to fine-tune an external control parameter, such as temperature, pressure or magnetic field, in contrast to the phenomena described above which occur universally without any fine-tuning. The explanation is that open extended dissipative dynamical systems i.e., systems not in thermodynamic equilibrium may go automatically to the critical state as long as they are driven slowly: the critical state is self-organized[87,95,99].

Fluctuations in time of atmospheric flows, as recorded by meteorological parameters such as pressure, temperature, wind speed etc., exhibit self - similar fluctuations in time , namely, a zigzag pattern of increase (decrease) followed by a decrease (increase) on time scales from seconds to years. Such jagged pattern for atmospheric variability (temporal) resembles the self - similar coastline structure. Long - range correlations in space and time, namely self - similar (fractal) fluctuations in space and time implies that the magnitude of the fluctuation (spatial or temporal) at any scale is related to other scales (larger and smaller) by a single parameter, namely, the scale factor which is a dimensionless number. Therefore, dynamical laws which govern the space - time fluctuations of smallest scale (turbulence, millimeters - seconds) fluctuations in space - time also apply for the largest scale (planetary, thousands of kilometers - years) in atmospheric flows throughout the globe. The co - operative existence of fluctuations of all scales gives rise to self - similar (coherent) space - time structures. The formation of such coherent structures which function as a unified whole has a special significance in the field of *Biology*, in the functioning of living systems. Pattern formation[100], i.e. morphogenesis, forms an integral part of *Life Sciences* and the vast amount of knowledge gathered in this field can beneficially be applied to other fields of science since self - similar space-time patterns are generic to nature, in particular, weather and climate patterns in meteorology. The multidisciplinery approach to the study of self-organized criticality will result in immense benefit to the scientific community as a whole in terms of transfer of new insights and techniques from one field to another.

## 2.10 Turbulent (chaotic) fluctuations and self-similar structure formation

The first phenomenological treatments of morphogenesis were built for fluid dynamics through the mathematical modelling of instabilities as those named after Faraday, Rayleigh and Benard, Rayleigh and Taylor, Kelvin and Helmholtz etc.[101],. Biological auto-organization and pattern formation have been studied over the past 40 years as non-equilibrium thermodynamic phenomena[102]. Biological systems exhibit high degree of co-operation in the form of long-range communication. The concept of co-operative existence of fluctuations in the organization of coherent structures have been identified as selforganized co-operative phenomena[103]. The study of the spontaneous, i.e., self - organized formation of structures in systems far from thermal equilibrium in open systems belongs to the multidisciplinary field of synergetics[104,105]. Plant kingdom exhibits examples of the most striking self - similar geometrical patterns[11] signifying *self - organized criticality* in the spatial structure formation.

Formation of structure begins by aggregation of molecules in a turbulent fluid (gas or liquid) medium. Turbulent fluctuations are therefore not dissipative, but serve to assemble and form coherent structures[106-108], for example, the formation of clouds in turbulent atmospheric flows[109]. Traditionally, turbulence is considered dissipative and disorganized. Yet, coherent (organized) vortex roll circulations (vortices) are ubiquitous to turbulent fluid flows[110,111]. The exact physical mechanism for the formation and maintenance of coherent structures, namely vortices or large eddy circulations in turbulent fluid flows is not yet identified. The most intense weather systems such as hurricanes have vividly spiraling cloud formation while the destructive tornado has spiraling (vortex) air flow in narrow funnel-like protuberances which reach down to earth and create devastating damage.

Recent studies show that clouds of all sizes[112] are self - similar in shape, which is consistent with commonly visualized shape of clouds as billows upon billows. Incidentally, it may be mentioned that cumulus clouds bear a close resemblance to cauliflowers. Meteorological textbooks commonly describe the cumulus clouds to have cauliflower - like structure. In the midst of turbulence in air flows, clouds retain their identity in shape and the most astonishing of all is the formation of ice crystals with exquisitely symmetrical structure.

Nature abounds in symmetrical structures from the macro - to the microscopic scales[113]. Perfect order appears to underlie apparent chaos in turbulent flows. Turbulence, namely, seemingly random fluctuations of all scales, therefore, plays a key role in the formation of self-similar coherent structures in the atmosphere. Such a concept is contrary to the traditional view that turbulence is dissipative, i.e. ordered growth of coherent form is not possible in turbulent flows. Selvam[69], Selvam and Fadnavis[114], Selvam *et al*[115], Joshi and Selvam[116] have shown that turbulent fluctuations self-organize to form self - similar structures in fluid flows.

Ramified branching networks serve to connect and assist in the functioning as a unified whole of self-similar *fractal* objects. A *fractal* object can be resolved into smaller interconnected component parts, which resemble the whole in shape. The self-similar architecture for *fractal* objects serves for collection and distribution of information/energy between the largest and smallest scales. For example, the river system collects water from tributaries, the lung architecture enables efficiency of oxygen absorption from air in the alveoli (the smallest branching structure). Jean[11] has emphasized the functional importance of ramified structures underlying self-similar *fractals* and gives reference to earlier studies, which show that such branching structures can be organized into hierarchies, which incorporate the *Fibonacci* mathematical sequence.

### 2.11 Self - similarity: A signature of identical iterative growth process

Self-similarity underlies all growth processes in nature. Jean[11] has emphasized the self-similar geometry of botanical elements. Self-similar structures are generated by iteration (repetition) of simple rules for growth processes on all scales of space and time. Such iterative processes are simulated mathematically by numerical computations such as $X_{n+1} = F(X_n)$ where $X_{n+1}$, the value of the variable X at $(n+1)^{th}$ computational step is a function F of its earlier value $X_n$. Mathematical models of real world dynamical systems are basically such iterative computational schemes implemented on finite precision digital computers. Computer precision imposes a limit (finite precision) on the accuracy (number of decimals) for numerical representation of X. Since X is a real number (infinite number of decimals) finite precision introduces round - off error in iterative computations from the first stage of computation. The model iterative dynamical system therefore incorporates round - off error growth. Computed growth patterns exhibit self-similar *fractal* structure which incorporate the *golden mean*[117]. The new science of *nonlinear dynamics and chaos* seeks to understand the physics of such self-similar patterns in computed and real world dynamical systems.

## 3. Fractals and Self-Organized Criticality in Atmospheric Physics

Fluid flows such as river flows, atmospheric flows, etc. are characterized by turbulence, namely, seemingly random fluctuations on all space and time scales. Traditionally, turbulence is considered dissipative and disorganized. Yet, coherent (organized) vortex roll circulations (vortices) are ubiquitous to turbulent fluid flows[109-111]. The exact physical mechanism for the formation and maintenance of coherent structures, namely vortices or large eddy circulations in turbulent fluid flows is not yet identified. Paradoxically, the more severe the turbulence (disorder), the more vivid the organization of coherent spiraling structures, e.g., the spiral cloud bands, and airflow circulations in the destructive hurricanes and tornadoes. Lovejoy and Schertzer[118] have shown conclusively that the seemingly irregular fluctuations of meteorological parameters are self-similar *fractals*, the power spectra exhibiting inverse power law form $f^{-\alpha}$ where $f$ is the frequency and the exponent $\alpha$ is different for different scale ranges. The fluctuations exhibit scale invariance or self-similarity for scale ranges with constant scale factor $\alpha$.

Nonlinear dynamical systems in nature such as atmospheric flows exhibit complex spatial patterns, e.g., cloud geometry, that lack a characteristic (single) length scale concomitant with temporal fluctuations that lack a single time scale. Objects in nature are in general *multifractals*, i.e. the *fractal dimension* is different for different scale ranges. The concept of *fractal dimension* introduced by Mandelbrot[6] provides a powerful tool for quantitative description of nonlinear fluctuations in real world and computed dynamical systems. Self-similar fractal structures are also found in finite precision computer realization of nonlinear mathematical models of dynamical systems. Self - similarity implies long - range spatiotemporal correlations. Lorenz[119] was the first to identify the sensitive dependence on initial conditions characteristic of such self - similar structures in finite precision computer realization of a simple model of atmospheric flows. Sensitive dependence on initial conditions precludes exact prediction and therefore named *deterministic chaos*[16] since deterministic model equations give chaotic numerical solutions. Numerical models, even with only a few degrees of freedom resemble real world dynamical systems in generating irregular (complex) fluctuations. The concept of *fractals*, i.e. self - similar fluctuations implies long - range correlations in space and time. Long - range spatiotemporal correlations are ubiquitous to dynamical systems in nature and are identified as signatures of self - organized criticality[7,14,99]. The fractal structure of atmospheric flows in space and time has been identified and discussed in detail by Lovejoy and his group[112,120-123]. Fractals and multifractals characterize fluid turbulence[44], atmospheric flows being a representative example of fluid turbulence. Atmospheric flows therefore exhibit *self-organized criticality*. Standard meteorological theory cannot explain satisfactorily[112] the observed self-similar structures to spatiotemporal pattern of atmospheric flows ranging from the turbulence (millimeters - seconds) to climatological (kilometers - years) scales.

The cooperative existence of fluctuations ranging in size - duration from a few millimeters - seconds (turbulence scale) to thousands of kilometers - years (planetary scale) result in the observed long - range spatiotemporal correlations, namely, fractal geometry to the global cloud cover pattern concomitant with inverse power law form for power spectra of temporal fluctuations documented by Lovejoy and Schertzer[120,121] and Tessier et al[112]. Long-range spatiotemporal correlations are ubiquitous to real world dynamical systems and are recently identified as signatures of self - organized criticality[7]. The physics of self - organized criticality is not yet identified. It is important to quantify the total pattern of fluctuations in atmospheric flows for predictability studies. Traditional numerical weather prediction models based on Newtonian continuum dynamics are nonlinear and require numerical solutions incorporating numerical integration schemes which are basically iterative computations. Finite precision computer realizations of such nonlinear models are sensitively dependent on initial conditions, now identified as deterministic chaos[16] resulting in unrealistic solutions. The physics of deterministic chaos is not yet identified. Selvam[124] has shown that round - off error approximately doubles on an average for each step of finite precision numerical iteration. Such round - off error doubling results in unrealistic solutions for numerical weather prediction and climate models which incorporate long - term numerical integration schemes with thousands of such iterations. Realistic modeling of atmospheric flows therefore requires alternative concepts for fluid flows and robust computational techniques which do not require round - off error prone calculus - based long - term numerical integration schemes.

### 3.1 Sensitivity to initial conditions in real world and model dynamical systems

The measured mathematical descriptors of physical stochastic processes which are nonstationary such as turbulent atmospheric flows exhibit sensitivity to initial conditions manifested as the widely documented selfsimilar growth of space-time structures which exhibit longe-range correlations between the small and large scales, a striking example being the devastating global scale anomalous weather events triggered by local perturbation such as the ElNino event. A general systems theory for fractal space-time fluctuations proposed by Selvam[69,125] visualizes the large scale as the envelope enclosing smaller scale fluctuations thereby accounting for the observed selfsimilar space-time structures ubiquitous to dynamical systems in nature.

Sensitivity to initial conditions in deterministic nonlinear models (e.g., temporal chaos in a finite set of ordinary differential equations like the Lorenz system or spatio-temporal chaos in an initial-boundary-value problem with one or more partial differential equations such as the Navier-Stokes equations describing turbulence) is due to the inevitable error (input plus finite precision round-off) feedback with amplification into the main stream computation in numerical integration schemes such as Runge-Kutta method used in computed solutions[69,125].

### 3.2 Observed structure of atmospheric flows and signatures of deterministic chaos

Recent advances in remote sensing and *in situ* measurement techniques have enabled atmospheric scientists to document the following new observational characteristics of turbulent shear flows in the planetary atmospheric boundary layer (ABL) where weather activity occurs. The ABL extends to about 10 km above the surface of the earth.

(*i*) The atmospheric flow consists of a full continuum of fluctuations ranging in size from the turbulence scale of a few millimetres to the planetary scale of thousands of kilometres.

(*ii*) The atmospheric eddy energy spectrum follows an inverse power-law of form $f^\alpha$ where $f$ is the frequency and $\alpha$ the exponent. The exponential power-law form for the eddy energy spectrum indicates self-similarity and scale invariance. The exponent $\alpha$ is found to be equal to 1.8 for both meteorological (time period in days) and climatological (time period in years) scales, which indicates a close coupling between the two scales[120,121,126-131].

(*iii*) Satellite cloud-cover photographs give evidence for the existence of helical vortex-roll circulations (or large eddies) in the ABL as indicated by the organization of clouds in rows and (or) streets, meso - scale (up to 100km) cloud clusters (MCC), and spiral bands in synoptic scale weather systems[132].

(*iv*) The structure of atmospheric flows is invariably helical (curved) as manifested in the visible cloud patterns of weather systems, e.g., all basic meso - scale structures such as medium scale tornado generating storms, squall lines, hurricanes, etc[110], and in particular the super - cell storm[133].

(*v*) Atmospheric flows give an implicit indication of the upscale transfer of a certain amount of energy inserted at much smaller scales, thereby generating the observed helical fluctuations[110,133].

(*vi*) The global cloud-cover pattern exhibits self - similar fractal geometrical structure and is consistent with the observed scale invariance of the atmospheric eddy energy spectrum[120,121,134] (see characteristic (*ii*) above).

Atmospheric weather systems exist as coherent structures consisting of discrete cloud cells forming patterns of rows and (or) streets, mesoscale cloud clusters (MCC), and spiral bands. These patterns maintain their identity for the duration of their appreciable lifetimes in the apparently dissipative turbulent shear flows of the ABL[109]. The existence of coherent structures (seemingly systematic motion) in turbulent flows in general, has been well established during the last 30 - 40 years of research into turbulence. However, it is still debated whether these structures are the consequences of some kind of instabilities (such as shear or centrifugal instabilities), or whether they are manifestations of some intrinsic universal properties of any turbulent flow[110].

Lovejoy and Schertzer[120,121] have provided conclusive evidence for the signature of deterministic chaos in atmospheric flows, namely the fractal geometry of global cloud-cover pattern and the inverse power-law form $f^{\alpha}$ where $f$ is the frequency and $\alpha$ the exponent for the atmospheric eddy energy spectrum. Atmospheric teleconnections, such as the El Nino and (or) Southern Oscillation (ENSO) cycles in weather patterns, that are responsible for devastating changes in normal global weather regimes[135-137] are also manifestations of long-range correlations in regional weather activity.

Meteorologists have documented in detail the nonlinear variability of atmospheric flows, in particular the interannual variability, i.e., the year to year fluctuations in weather patterns. The interannual variability of atmospheric flows is nonlinear and exhibits fluctuations on all scales ranging up to the length of data period (time) investigated. The broadband spectrum of atmospheric interannual variability has embedded dominant quasiperiodicities such as the quasibiennial oscillation (QBO) and the ENSO (*El Nino/Southern Oscillation*) cycle of 3 to 7 years which are identified as major contributors to local climate variability, in particular, the monsoons which influence agriculture dependent world economies. ENSO is an irregular self - sustaining cycle of alternating warm and cool water episodes in the Pacific Ocean. Also called *El Nino - La Nina*, *La Nina* refers to the cool part of the weather cycle while *El Nino* is associated with a reversal of global climatic regimes resulting in anomalous floods and droughts throughout the globe. It is of importance to quantify the total pattern of fluctuations for predictability studies.

### 3.3 Limitations of conventional atmospheric boundary layer (ABL) models

Presently available models for ABL turbulent flows are incapable of identifying the coherent helical structural form intrinsic to turbulence. Also, the models do not give realistic simulations of the space-time averages for the thermodynamic parameters and the fluxes of buoyant energy, mass, and momentum because of the following inherent limitations.

(*i*) The physics of the observed coherent helical geometric structure inherent in turbulent flows is not yet identified, and therefore the structural form of turbulent flows cannot be modelled.

(*ii*) By convention, the Newtonian continuum dynamics of the atmospheric flows are simulated by the Navier Stokes (NS) equations, which are inherently nonlinear, and being sensitive to initial conditions, give chaotic solutions characteristic of deterministic chaos.

(*iii*) The governing equations do not incorporate the mutual coexistence and interaction of the full spectrum of atmospheric fluctuations that form an integral part of atmospheric flows[128,138-140].

(*iv*) The limitations of available computer capacity necessitate severe truncations of the governing equations, thereby generating errors of approximations.

(*v*) The above-mentioned uncertainties are further magnified exponentially with time by computer round-off errors and result in unrealistic solutions[141,142]. Recent exhaustive studies by Weil[142] and others also indicate that existing numerical models of atmospheric boundary layer flows require major revisions to incorporate an understanding of turbulence and diffusion in boundary layer flows. Recently, there has been growing conviction that current numerical weather prediction models are inadequate for accurate forecasts[118,144-148]. Numerical modelling of atmospheric flows, diffusion, and cloud growth therefore require alternative concepts and computational techniques.

### 3.4 Traditional numerical weather prediction, deterministic chaos and predictability

Standard models for turbulent fluid flows in meteorological theory cannot explain satisfactorily the observed *multifractal* (space - time) structures in atmospheric flows[112,149]. Traditionally, meteorological theory is based on the following concepts. The turbulent atmospheric flows are governed by the mutual interaction of a large number of factors, i.e. variables such as pressure, temperature, moisture content, wind speed, etc. Historically, Richardson[150], in 1922, formulated quantitative computational method for weather prediction as follows. The prediction of future flow pattern is based on mathematical equations for the rate of change $dX/dt$ of component variable $X$ with time $t$. The rate of change with time $dX/dt$ of any variable $X$ is generally a nonlinear function of all the other interacting variables and therefore analytical solution for $X$ is not available. The evaluation of any variable $X$ with time is then computed numerically from the iterative equation

$$X_{n+1} = X_n + \left(\frac{dX}{dt}\right)_n dt$$

where the subscript $n$ denotes the time step and the rate of change $dX/dt$ is assumed to be continuous for small changes in time $dt$, an assumption based on *Newtonian* continuum dynamics. The successive values of $X$ are then computed iteratively, a process known as numerical integration. The prediction equation for the variable $X$ has intrinsic error feedback loop since the value of $X$ at each step is a function of its earlier value in such numerical integration computational techniques. The fundamental (basic) error in numerical computations is the round - off error of finite precision computations. Blank[151] mentions that when solving differential or other dynamical systems on a computer, the effects of finiteness (round-off) can sometimes be very drastic. When we work with fixed precision system, not all real numbers are even representable and arithmetic does not have the properties that we are used to[152,153].

Lorenz[154] has discussed chaotic behaviour when continuum equations are solved numerically as difference equations. Climate modelling concepts has come under criticism lately since uncertainty in input parameter values can give drastically different results[155]. Selvam[124] has shown that round - off error approximately doubles on an average at each step of iteration. Such error doubling at each step in numerical integration will result in the round-off error propagating into the mainstream (digits place and above) computation within 50 iterations using single precision (7th decimal place accuracy) digital computers. In addition, any uncertainties in specifying the initial value of the variable X will also grow exponentially with time and give unrealistic solutions. Numerical solutions are therefore sensitively dependent on initial conditions. Deterministic governing equations, namely evolution equations which are precisely defined and mathematically formulated give chaotic solutions because of sensitive dependence on initial conditions. Finite precision computer realizations of nonlinear mathematical models of dynamical systems therefore exhibit deterministic chaos. Computed model solutions are therefore mere mathematical artifacts of the universal process of round-off error growth in iterative computations[124] and the computed domain is the successive cumulative integration of round - off error domains analogous to the formation of large eddy domains as envelopes enclosing turbulent eddy fluctuation domains such as in atmospheric flows[69,125,156-158].

Computed solutions, therefore qualitatively resemble real world dynamical systems such as atmospheric flows with manifestation of *self - organized criticality. Self - organized criticality*, i.e. long - range spatiotemporal correlations, originates with the primary perturbation domains corresponding respectively to round-off error and dominant turbulent eddy fluctuations in model and real world dynamical systems. Computed solutions, therefore, are not true solutions. The vast body of literature investigating chaotic trajectories in recent years (since 1980) document, only the round - off error structure in finite precision computations. Stewart[159] mentions that in the absence of analytical (true) solutions the fidelity of computed solutions is questionable.

Hacker *et al*[160] discuss predictability of dynamical systems in general and atmospheric flows in particular as follows. The study of predictability is multifaceted and appears in diverse fields. For purposes of discussion following the lead of Joseph Tribbia[161], Hacker *et al.*[160] adopted the definition of predictability proposed by Thompson[162] which is "the extent to which it is possible to predict (the-atmosphere) with a theoretically complete knowledge of the physical laws governing it". More precisely, Hacker *et al*[160] interpreted this as the state-dependent rate of divergence of trajectories in phase space given complete knowledge of system dynamics. Therefore, predictability is intrinsic to a system, and the atmosphere (most likely) has predictability properties distinct from those of any model. Similar statements can be made about biological and all other dynamical systems. We can exactly describe and solve for the evolution of some simple systems analytically, but we are faced with the frustrating reality that we cannot precisely know the predictability of more complex-systems. Thus, much of our science is the pursuit of an unknowable goal[160].

Zupanski and Navon[163] state that uncertainty estimation is becoming an important new research discipline, crosscutting many scientific areas. The mathematical concept of uncertainty estimation is based on probability theory and statistics, estimation theory, information theory, and control theory. Theoretical aspects of uncertainty estimation are generally well understood for linear models (operators) and Gaussian distribution. In the geosciences, however, nonlinear models are typically used; thus, the Gaussian probability assumption may not be the best option. In addition, models of geosciences systems are typically high-dimensional, with state variable dimensions of the order of $10^6$–$10^7$. At the same time, the mathematical concept of uncertainty estimation, algorithmically defined by smoothing and/or filtering, is relatively simple, and a common mathematical framework can be applied across disciplines. These facts create a challenging problem for uncertainty estimation, requiring new scientific developments and cross-disciplinary efforts.

Historically, deterministic chaos, the origin of uncertainty in computed solutions was identified nearly a century ago by *Poincare* in his study of the three body problem[164]. Lack of high speed computational machines precluded exhaustive studies of nonlinear behavior and approximate linearized solutions of nonlinear systems alone were studied. With the advent of electronic digital computers in late 1950s, Lorenz[119] identified deterministic chaos in a simple model of atmospheric flows. *Lorenz*'s result captured the attention of scientists in all branches of science since a simple set of equations exhibits chaotic behaviour similar to the complex, irregular (unpredictable) fluctuations exhibited by real world dynamical systems. Till then it was believed that complex behavior results from complexity in the governing parameters and the mathematical formulations. *Lorenz*'s model demonstrated that simple models can demonstrate complex behavior of real world dynamical systems.

The computed trajectory is plotted graphically in phase space of dimension *m* where *m* is the number of variables representing the dynamical system. For example, a particle in motion can be represented completely at any instant by its position and momenta in the *x, y and z* directions, i.e., 6-dimensional phase space. The line joining the successive points in time gives the trajectory of the particle in phase space. The trajectory traces the *strange attractor*, so named because of its strange convoluted shape being the final destination of all trajectories in the phase space. Two trajectories, initially close together diverge exponentially with time though still within the *strange attractor* domain, thereby exhibiting sensitive dependence on initial conditions or *deterministic chaos*. The *strange attractor* exhibits self-similar *fractal* geometry similar to the space-time *fractal* structure or *self-organized criticality* exhibited by real world dynamical systems. Selvam[124] has shown that the *strange attractor* has the quasicrystalline structure of the quasiperiodic *Penrose* tiling pattern. There is a very close similarly between the geometrical patterns generated during iterative computations and those found in nature[117,165]. Iterative computations generate patterns strongly reminiscent of plant forms and clearly these curious configurations show that the rules responsible for the construction of elaborate living tissue structures could be absurdly simple[166]. In summary, self - similar space - time structures or *self - organized criticality* is ubiquitous to dynamical systems in nature and also to mathematical models of dynamical systems which incorporate finite precision iterative computations with resultant feedback and magnification of round - off error primarily, in addition to initial errors. Iterative computations result in the cumulative addition (integration) of the progressively increasing round - off error. Persistent perturbations, though small in magnitude are therefore capable of generating complex space - time structures with *fractal* self - similar geometry because of feedback with amplification.

### 3.5 Current techniques in numerical weather prediction (NWP): major drawbacks

Present day weather/climate predictions are probabilistic ensemble forecasts. Atmospheric evolution is chaotic, i.e. sensitive to initial - condition uncertainty. However, with modern-day supercomputers, we can run weather forecast models many times from very slightly different initial conditions, consistent with the uncertainties to estimate the effect of this initial - condition uncertainty. The resulting forecasts can be combined to produce a forecast probability distribution and is basically an ensemble weather forecast[167].

Roebber and Tsonis[168] describe the ensemble forecasting method and its drawbacks as follows. Ensemble prediction has become an indispensable tool in weather forecasting. One of the issues in ensemble prediction is that, regardless of the method, the prediction error does not map well to the underlying physics (i.e., error estimates do not project strongly onto physical structures). The fundamental problem of weather forecasting is to identify the range of possible meteorological scenarios that might evolve from a given initial state, and determine whether multiple solutions have high probability (low confidence in an individual solution) or if a single evolution is the most likely evolution (high confidence). This probabilistic view is necessitated by the complexity of atmospheric dynamics (e.g., Lorenz[119]; see a general review in Kalnay[169]). Specifically, limits to deterministic predictability originate from two general sources: model error and initial condition error. Model error arises because of imperfections in our characterizations of the laws of nature, arising either through parameterization of complex and/or poorly understood physics (such as boundary layer and cloud microphysical processes) or an inability to resolve atmospheric processes smaller than a certain threshold (e.g., atmospheric convection with a 10-km gridpoint model), with resultant upscale error growth. Initial condition error arises because of the finite spatial availability of observed data (including some variables that are not observed at all), missing data, inaccuracies in the data, and imperfect analysis techniques. All of these errors, even with a perfect model, will grow nonlinearly over time, eventually swamping the forecast signal[119,170,171]. The rate of this error growth and hence the lead time at which predictability is lost depends on the stability of the evolving flow[170], which in addition is affected by factors such as the large-scale flow pattern, season, and geographical domain[172,173]. Ensemble forecast systems have been developed as a means to quantify forecast uncertainty, using a variety of methods to simulate analysis and model uncertainties[168].

Lovejoy and Schertzer[118] have summarized the current status of NWP as follows. Twenty years ago the goal of weather forecasting was to determine the (supposedly unique) state of the atmosphere at some time in the future, whereas today, ensemble forecasting systems have instead the goal of determining the possible states of tomorrow's weather including their probabilities of occurrence. This new goal therefore corresponds to a transition from deterministic to stochastic forecasts. Today's ensemble forecasting systems therefore require knowledge of the underlying stochastic structure of the deterministic equations.

The current ensemble forecasting technique is essentially a stochastic—deterministic hybrid which is indirect and problematic on several counts. The main difficulties are (i) that it is based on a deterministic framework for the initial objective analysis—which uses statistics only to describe measurement errors — and not the fields themselves — and (ii) which assumes that the fields evolve according to deterministic nonlinear partial differential equations. While deterministic assumptions may be appropriate for descriptions and models at the dissipation scale, stochastic ones are more appropriate at lower space–time resolution (if only because an infinite number of different dissipation scale fields give rise to the same low resolution analysis fields).

Lovejoy and Schertzer[118] have discussed the issue of distinguishing natural from anthropogenic variability and the problem of outliers as follows. Conclusions about anthropogenic influences on the atmosphere can only be drawn with respect to the null hypothesis, i.e. one requires a theory of the natural variability, including knowledge of the probabilities of the extremes at various resolutions. At present, the null hypotheses are classical so that they assume there are no long range statistical dependencies and that the probabilities are thin-tailed (i.e. exponential). However it is seen that cascades involve long range dependencies and (typically) have fat tailed (algebraic) distributions in which extreme events occur much more frequently and can persist for much longer than classical theory would allow. Indeed, the problem of statistical "outliers" may generally be a consequence of the failure of highly variable cascade data to fit into relatively homogeneous, regular, classical geostatistical frameworks.

## 4. Applications of Nonlinear Dynamics and Chaos for Weather Prediction: Current Status

At present, the signatures of deterministic chaos, namely the fractal geometrical structure concomitant with $1/f$ noise, have been conclusively identified in model and real world atmospheric flows, and the fractal dimension of the strange attractor traced by atmospheric flows has been estimated with recently developed numerical algorithms[174,175], which use the time series data of meteorological parameters, e.g., rainfall, temperature, wind speed, etc. However, such estimations of the fractal dimension have not helped resolve the problem of the formulation of a simple closed set of governing equations for atmospheric flows[176-179] mainly because the basic physics of deterministic chaos is not yet identified. A complete review of applications of concepts in *nonlinear dynamics and chaos* in atmospheric sciences has been given by Zeng *et al*[180].

During the past three decades, Lovejoy and his group[118] have done extensive observational and theoretical studies of fractal nature of atmospheric flows and emphasize the urgent need to formulate and incorporate quantitative theoretical concepts of fractals in mainstream classical meteorological theory. The empirical analyses summarized by Lovejoy and Schertzer[118] directly demonstrate the strong scale dependencies of many atmospheric fields, showing that they depend in a power law manner on the space–time scales over which they are measured. In spite of intense efforts over more than 50 years, analytic approaches have been surprisingly ineffective at deducing the statistical properties of turbulence. Atmospheric science labors under the misapprehension that its basic science issues have long been settled and that its task is limited to the application of known laws — albeit helped by ever larger quantities of data themselves processed in evermore powerful computers and exploiting ever more sophisticated algorithms. Conclusions about anthropogenic influences on the atmosphere can only be drawn with respect to the null hypothesis, i.e. one requires a theory of the natural variability, including knowledge of the probabilities of the extremes at various resolutions. At present, the null hypotheses are classical so that they assume there are no long range statistical dependencies and that the probabilities are thin-tailed (i.e. exponential). However observations show that cascades involve long-range dependencies and (typically) have fat tailed (algebraic) distributions in which extreme events occur much more frequently and can persist for much longer than classical theory would allow[118].

Dessai and Walter[181] argue that there is enough evidence, to show that complexity and its theory of selforganized criticality (SOC) have considerable potential to increase our understanding of the atmospheric sciences and emphasized the urgent need to incorporate fundamental concepts of SOC in atmospheric as follows. Meteorologists and climatologists have largely ignored SOC. Although large power events are comparatively rare, events can and do happen on all scales, with no different mechanism needed to explain the rare large events than that which explains the smaller, more common ones[182]. In the atmospheric sciences there has been little application of what some have considered the leading candidate for a unified theory of complexity, namely, SOC. In his review of complexity and climate, Rind[183] concludes climate, like weather will likely always be complex: *"determinism in the midst of chaos, unpredictability in the midst of understanding."* Rind[183] warns that it is still not known if complexity is relevant to climate science. Theories of complexity, such as SOC, have been underrepresented in the atmospheric sciences because of their "soft science" character[181]. Atmospheric sciences have historically developed from centuries of advancement in the hard sciences, such as physics, mathematics and chemistry, etc. It would have been unlikely to see a quick transition from the classical reductionist and reproducible science approach towards an abstract, holistic and probabilistic complex science. Proof of this is the fact that only a small number of scientists have cited the few applications of these theories in the atmospheric sciences, e.g., Vattay and Harnos[184], Lovejoy and Schertzer[118] (and all references therein) conclude that the multifractal approach yields a convenient framework for the analysis and simulation of highly nonlinear meteorological fields over a wide range of scales and intensities and Selvam[69,125,156,157] has developed a cell dynamical system model (general systems theory) for self - organized criticality[181].

Tsonis *et al*[185] have recently applied the concept of networks for the observed scale - free pattern for atmospheric flows as follows. Advances in understanding coupling in complex networks offer new ways of studying the collective behavior of interactive systems and already have yielded new insights in many areas of science. From this initial application of networks to climate it appears that atmospheric fields can be thought of as a network of interacting points whose collective behavior may exhibit properties of small world networks. This ensures the efficient transfer of information. In addition, the scale-free architectures guarantee stability. Furthermore, supernodes in the network identify teleconnection patterns. As was demonstrated in Tsonis[186], these teleconnections are not static phenomena, but their spatiotemporal variability is affected by large (global) changes. Complex networks describe many natural and social dynamical systems, and their study has revealed interesting mechanisms underlying their function. The novelty of networks is that they bring out topological/geometrical aspects that are related to the physics of the dynamical system in question, thus providing a new and innovative way to treat and investigate nonlinear systems and data. While several advances have been made, this area is still young and the future is wide open. This introductory paper presented some fundamental aspects of networks and some preliminary results of the application of networks to climatic data, which indicate that networks delineate some key features of the climate system. This suggests that networks have the potential to become a new and useful tool in climate research[185].

**4.1 Space– time cascade model for fractal fluctuations in atmospheric flows**

Lovejoy and Schertzer[118], the pioneers in the study of nonlinear dynamical characteristics of atmospheric flows have proposed a space - time cascade model for a realistic simulation of weather and climate as summarized in the following. In spite of the unprecedented quantity and quality of meteorological data and numerical models, there is still no consensus about the atmosphere's elementary statistical properties as functions of scale in either time or in space. The proposed model is a new synthesis based on a) advances in the last 25 years in nonlinear dynamics, b) a critical re-analysis of empirical aircraft and vertical sonde data, c) the systematic scale by scale, space–time exploitation of high resolution remotely sensed data and d) the systematic re-analysis of the outputs of numerical models of the atmosphere including reanalyses, e) a new turbulent model for the emergence of the climate from "weather" and climate variability. Lovejoy and Schertzer[118] conclude that Richardson's old idea of scale by scale simplicity — today embodied in multiplicative cascades — can accurately explain the statistical properties of the atmosphere and its models over most of the meteorologically significant range of scales, as well as at least some of the climate range[187]. The resulting space– time cascade model combines these nonlinear developments with modern statistical analyses; it is based on strongly anisotropic and intermittent generalizations of the classical turbulence laws of Kolmogorov, Corrsin, Obukhov, and Bolgiano.

Lovejoy and Schertzer[118] have given an overview of a body of work carried out over the last 25 years aiming at a scale by scale understanding of the space–time statistical structure of the atmosphere and its models. The proposed new synthesis would not be possible without technologically driven revolutions in both data quantity and quality as well as in numerical modeling and data processing. Also key for this synthesis are advances in our understanding of nonlinear dynamics (especially cascades, multifractals, and their anisotropic extensions), and in the corresponding data analysis techniques. Although there are many gaps to fill, it is remarkable that a relatively simple picture of the atmosphere as a system of interacting anisotropic cascades seems to be consistent with some of the largest and highest quality satellite, lidar, dropsonde and aircraft campaigns to date collectively measuring passive and active radiances over the long and short wave regimes, as well as in situ wind, temperature, humidity, potential temperature, pressure and other variables. It also holds remarkably well for reanalyses and other numerical models of the atmosphere. It leads to a natural distinction between the weather and climate and predicts the transition to the climate at approximately equal to 10 days as a dimensional transition from a weather system (where both long - range space and time correlations are important) to a climate system dominated by long - range temporal correlations.

Lovejoy and Schertzer[118] conclude that, in any case, some coherent picture is urgently needed to replace the aging and untenable (but still dominant!) 2D isotropic/3D isotropic turbulence model.

## 4.2 General systems theory for fractal space-time fluctuations in atmospheric flows

Selvam[69,125,156-158] has recently developed general systems theory for fractal space-time fluctuations based on the concept that the larger scale fluctuation can be visualized to emerge from the space-time averaging of enclosed small scale fluctuations, thereby generating a hierarchy of self-similar fluctuations manifested as the observed eddy continuum in power spectral analyses of fractal fluctuations. Such a concept results in inverse power law form incorporating the golden mean $\tau$ for the space-time fluctuation pattern and also for the power spectra of the fluctuations. The predicted distribution is close to the Gaussian distribution for small-scale fluctuations, but exhibits *fat long tail* for large-scale fluctuations. The general systems theory, originally developed for turbulent fluid flows, provides universal quantification of physics underlying fractal fluctuations and is applicable to all dynamical systems in nature independent of its physical, chemical, electrical, or any other intrinsic characteristic.

Macroscale coherent structures emerge by space-time integration of microscopic domain fluctuations in fluid flows. Such a concept of the autonomous growth of atmospheric eddy continuum with ordered energy flow between the scales is analogous to Prigogine's[107] concept of the spontaneous emergence of order and organization out of apparent disorder and chaos through a process of self-organization.

The problem of emergence of macroscopic variables out of microscopic dynamics is of crucial relevance in biology[188]. Biological systems rely on a combination of network and the specific elements involved[76]. The notion that membership in a network could confer stability emerged from Ludwig von Bertalanffy's description of general systems theory in the 1930s and Norbert Wieners description of cybernetics in the 1940s. General systems theory focused in part on the notion of flow, postulating the existence and significance of flow equilibria. In contrast to Cannon's concept that mechanisms should yield homeostasis, general systems theory invited biologists to consider an alternative model of homeodynamics in which nonlinear, non-equilibrium processes could provide stability, if not constancy[189]. The cell dynamical system model for coherent pattern formation in turbulent flows[69,125,156] may provide a general systems theory for biological complexity. General systems theory is a logical mathematical field, the subject matter of which is the formulation and deduction of those principles which are valid for 'systems' in general, whatever the nature of their component elements or the relations or 'forces' between them[9,10,190].

In summary, the model predicts the following: (i) The eddy continuum consists of an overall logarithmic spiral trajectory with the quasiperiodic *Penrose* tiling pattern for the internal structure. (ii)The successively larger eddy space-time scales follow the Fibonacci number series. (iii) The probability distribution $P$ of fractal domains for the $n^{th}$ step of eddy growth is equal to $\tau^{-4n}$ where $\tau$ is the golden mean equal to $(1+\sqrt{5})/2$ ($\approx 1.618$). The probability distribution $P$ is close to the statistical normal distribution for $n$ values less than 2 and greater than normal distribution for $n$ more than 2, thereby giving a *fat, long tail*. (iv) The probability distribution $P$ also represents the relative eddy energy flux in the fractal domain. The square of the eddy amplitude (variance) represents the eddy energy and therefore the eddy probability density $P$. Such a result that the additive amplitudes of eddies when squared represent probabilities, is exhibited by the sub-atomic dynamics of quantum systems such as the electron or proton[191-193]. Therefore fractal fluctuations are signatures of quantum - like chaos in dynamical systems. (v) The universal algorithm for self - organized criticality is expressed in terms of the universal *Feigenbaum*'s constants[194] *a* and *d* as $2a^2 = \pi d$ where the fractional volume intermittency of occurrence $\pi d$ contributes to the total variance $2a^2$ of fractal structures. (vi) The *Feigenbaum*'s constants are expressed as functions of the *golden mean*. The probability distribution *P* of fractal domains is also expressed in terms of the Feigenbaum's constants *a* and *d*.

The model predicted inverse power law distribution has been identified in time series of meteorological parameters[114-116,158].

## 5. Mathematical and physical topics relating chaos theory applied to atmosphere sciences

- **Spatio-temporal chaos in spatially extended oscillatory systems**

Sherratta et al[195] have applied the concept of spatiotemporal chaos considering a classic example from ecology: wavetrains in the wake of the invasion of a prey population by predators. In systems with cyclic dynamics, invasions often generate periodic spatiotemporal oscillations, which undergo a subsequent transition to chaos. The periodic oscillations have the form of a wavetrain and occur in a band of constant width. In applications, a key question is whether one expects spatiotemporal data to be dominated by regular or irregular oscillations or to involve a significant proportion of both. This depends on the width of the wavetrain band. Wavetrains are a fundamental solution type in spatially extended oscillatory systems. They were observed in the Belousov–Zhabotinskii reaction more than 30 years ago, and experimental and theoretical demonstrations of their importance are now widespread, in applications including hydrodynamics, solar cycles, chemical reactions, cell biology, and ecology. Moreover, wavetrains provide the background state for many more complicated behaviors, including spatiotemporal chaos[195].

- **Stochastic resonance for millennial-scale oscillations in climate**

Reconstructions of oxygen isotopes ($\delta^{18}O$) from Greenland ice cores have revealed that the last glacial period was interrupted by rapid climate swings in the Northern Hemisphere, with an approximate spacing of 1500 yr. These Dansgaard– Oeschger (DO) stadial–interstadial (cold–warm) transitions were accompanied by increases in surface air temperature ranging from $9^o$ to $16^oC$ over Greenland (completed in years to decades), followed by a long-lasting cooling. Although the volume of proxy data documenting these millennial-scale oscillations is continuously growing, the physical mechanism responsible for their existence and the causes of the major reorganization of the ocean–atmosphere coupled system remain poorly understood.

Stochastic resonance, a process that results from a combination of a weak (subthreshold) periodic forcing and noise, was shown to produce stadial– interstadial transitions whose characteristics share many similarities with the DO. In addition, the statistical properties of the climate shifts generated through this process appear to be consistent with those inferred from the Greenland Ice Core Project (GRIP) climate record. The stochastic resonance mechanism requires a bistable system; that is, a nonlinear system with two stable states separated by a potential barrier. While the weak periodic forcing modulates the height of the barrier, the noise allows transitions from one state to another. The combination of both processes results in an oscillation that synchronizes at a particular phase of the forcing. Coherence resonance (or autonomous stochastic resonance), however, can arise in more general excitable systems without the need for external periodic forcing. It results from the combination of noise and a periodic limit cycle internal to the climate system itself, with the noise level controlling the periodicity[196].

- **Stochastic resonance in extended systems**

  Wio et al[197] state that *Stochastic resonance* (SR) is nowadays a paradigm of the constructive effects of fluctuations on nonlinear systems. Since its discovery a quarter of century ago − and besides exploring related phenomena like e.g., *coherence resonance* − interest on SR has gradually shifted towards increasingly complex systems, *networks* and nonlinear *media*. Wio et al[197] have investigated the interplay between chaos and noise on an extended chaotic system, namely a spatiotemporally chaotic "toy model", useful for climate research, analyzing the interplay between the *deterministic noise* and a real random process. The results show clear numerical evidence of *two* SRlike behaviors. On one hand, a "normal" SR phenomenon occurring at frequencies that correspond to a system's periodic behavior. On the other hand a SSSR (*systemsize stochastic-resonance*)-like behavior, indicating that there is an optimal system *size* for the spatial system's response. The effect of the noise is stronger when chaos is underdeveloped. When the system is also subject to an external periodic signal, it reveals to be only weakly sensitive to the presence of noise and of external forcing. The above indicated results are of relevance for forecasting. The noise is too weak to produce changes in the spatial structure or to alter the main frequencies, but it is much stronger than the "self-generated" deterministic noise. The main conclusion is that forecasting can be improved at the resonant frequencies by the presence of external noise, due to the suppression of the self-generated chaotic noise[197].

- **Upwelling and Internal tides on the continental shelf**

  Internal tides on the continental shelf can be intermittent as a result of changing hydrographic conditions associated with wind-driven upwelling. In turn, the internal tide can affect transports associated with upwelling. Internal waves cause high-frequency variability in the turbulent kinetic energy in both the bottom and surface boundary layers, causing periodic restratification of the inner shelf in the area of the upwelling front. Increased vertical shear in the horizontal velocity resulting from the superposition of the upwelling jet and the internal tide results in intermittent patches of intensified turbulence in the mid–water column[198].

- **Analysis of ENSO using stochastic theories**

  The ElNino/SouthernOscillation (ENSO) phenomenon is the dominant climatic fluctuation on interannual timescales. It is an irregular oscillation with a distinctive broadband spectrum. On time scales ranging from the annual to the decadal, the dominant form of variability within the climate system is the El Nino/Southern Oscillation (ENSO) phenomenon whose centre of action is in the equatorial central and eastern Pacific. The variability is highly coherent spatially in both ocean and atmosphere and tends to be dominated by a relatively small number of large scale patterns in both media. The first theory posits that the irregularity derives primarily from a chaotic interaction of various slow modes and therefore assumes implicitly that the fast modes are of secondary importance. The second theory contends instead that the slow time-scale modes are not chaotic but are instead 'disrupted' in a stochastic fashion by the fast modes. So the area of ENSO irregularity is still not a settled question[199].

- **Nonstationary quasiperiodic statistics**

  Most climate fluctuations may be modulated by a variety of periodic or quasi-periodic deterministic forcing (e.g. diurnal, seasonal or Milankovitch cycles). These process modulations often induce cyclostationary (CS) behaviour defined as periodic correlations. It is proposed that quasi-periodic correlations may appear when the inertia (memory) of a process is modulating its pure CS property. This idea is reinforced by the CS observation, suggesting that ENSO could not only be modulated by each month of a season but also by several months together. The hypothesis made here is a kind of a CS that extends over time (called Extended CycloStationarity, ECS) and needs to develop a new kind of statistical tool to compute autocorrelation averaged over time[200].

- **Delay differential modeling for El Nino/Southern Oscillation (ENSO) variability**

  Zaliapin and Ghil[201] consider a highly idealized model for El Nino/Southern Oscillation (ENSO) variability, as introduced in an earlier paper. The model is governed by a delay differential equation for sea-surface temperature $T$ in the Tropical Pacific, and it combines two key mechanisms that participate in ENSO dynamics: delayed negative feedback and seasonal forcing. Zaliapin and Ghil[201] perform a theoretical and numerical study of the model in the three-dimensional space of its physically relevant parameters: propagation period $\tau$ of oceanic waves across the Tropical Pacific, atmosphere-ocean coupling $\kappa$, and strength of seasonal forcing $b$. Phase locking of model solutions to the periodic forcing is prevalent: the local maxima and minima of the solutions tend to occur at the same position within the seasonal cycle. Such phase locking is a key feature of the observed El Nino (warm) and La Nina (cold) events. The phasing of the extrema within the seasonal cycle depends sensitively on model parameters when forcing is weak. Zaliapin and Ghil[201] also study co-existence of multiple solutions for fixed model parameters and describe the basins of attraction of the stable solutions in a one-dimensional space of constant  initial model histories[201].

- **Climate tipping elements**

  The term ''tipping point'' commonly refers to a critical threshold at which a tiny perturbation can qualitatively alter the state or development of a system. Here the term ''tipping element'' is used to describe large-scale components of the Earth system that may pass a tipping point. Earth's history provides evidence of nonlinear switches in state or modes of variability of components of the climate system. Such past transitions may highlight potential tipping elements under anthropogenic forcing, but the boundary conditions under which they occurred were different from today, and anthropogenic forcing is generally more rapid and often different in pattern. Therefore, locating potential future tipping points requires some use of predictive models, in combination with paleodata and/or historical data. A variety of tipping elements could reach their critical point within this century under anthropogenic climate change. The greatest threats are tipping the Arctic sea-ice and the Greenland ice sheet, and at least five other elements could surprise us by exhibiting a nearby tipping point. This knowledge should influence climate policy[202].

- **Climate tipping as a noisy bifurcation**

  A major component of nonlinear dynamics is the theory of bifurcations, these being points in the slow evolution of a system at which qualitative changes or even sudden jumps of behaviour can occur. Tipping events, corresponding mathematically to dangerous bifurcations, pose a likely threat to the current state of the climate because they cause rapid and irreversible transitions. Also, there is evidence that tipping events have been the mechanism behind climate transitions of the past. There is currently much interest in examining climatic tipping points, to see if it is feasible to predict them in advance. Using techniques from bifurcation theory, recent work looks for a slowing down of the intrinsic transient responses, which is predicted to occur before an instability is encountered. This is done, for example, by determining the short-term autocorrelation coefficient in a sliding window of the time series: this stability coefficient should increase to unity at tipping. Such studies have been made both on climatic computer models and on real paleoclimate data preceding ancient tipping events. Model studies give hope that these tipping events are predictable using time series analysis: when applied to real geological data from past events prediction is often remarkably good but is not always reliable. Concerning the techniques of time-series analysis, two developments in related fields are of interest. First, theoretical physicists are actively developing methods of time-series analysis that take into account unknown nonlinearities, allowing for short term predictions even if the underlying deterministic system is chaotic. These methods permit, to a certain extent, the separation of the deterministic, chaotic, component of the time series from the noise. As several of the tipping events involve chaos, nonlinear time series analysis is a promising complement to the classical linear analysis[203].

- **Power-law persistence in the atmosphere: analysis and applications**

  Bunde et al[204] reviewed recent results on the appearance of long-term persistence in climatic records and their relevance for the evaluation of global climate models and rare events. The persistence can be characterized, for example, by the correlation $C(s)$ of temperature variations separated by $s$ days. Bunde et al204 showed that, $C(s)$ decays for large $s$ as a power law, $C(s) \sim s^{-\gamma}$. For continental stations, the exponent $\gamma$ is always close to 0.7, while for stations on islands $\gamma \approx 0.4$. In contrast to the temperature fluctuations, the fluctuations of the rainfall usually cannot be characterized by long-term power-law correlations but rather by pronounced short-term correlations. The universal persistence law for the temperature fluctuations on continental stations represents an ideal test-bed for the state of-the-art global climate models and allows us to evaluate their performance. In addition, the presence of long-term correlations leads to a novel approach for evaluating the statistics of rare events[205-207].

## 6. Conclusion

The summit statement of the climate prediction project[208] emphasizes the need for realistic climate/weather prediction as follows. Considerably improved predictions of the changes in the statistics of regional climate, especially of extreme events and high-impact weather, are required to assess the impacts of climate change and variations. Investing today in climate science will lead to significantly reduced costs of coping with the consequences of climate change tomorrow. Despite tremendous progress in climate modeling and the capability of high-end computers in the past 30 years, our ability to provide robust estimates of the risk to society, particularly from possible catastrophic changes in regional climate, is constrained by limitations in computer power and scientific understanding. To estimate the quality of a climate prediction requires an assessment of how accurately we know and understand the current state of natural climate variability, with which anthropogenic climate change interacts[208].

Numerical weather/climate prediction models do not give realistic forecasts[153] (see Sec. 3.4) because of the following inherent limitations: (1) the continuum dynamical system such as atmospheric flows is computed as a discrete dynamical system with implicit assumption of subgrid - scale homogeneity (2) model approximations and arbitrary assumptions (3) the governing equations do not incorporate the dynamical interactions and co - existence of the complete spectrum of turbulent fluctuations which form an integral part of the large coherent weather systems[118,120,121, 125,139,209,210] (4) binary number representation in digital computers precludes exact number representation at the data input stage itself (5) round - off error of finite precision computer arithmetic magnifies exponentially with time the above uncertainities and gives unrealistic solution[124,140]. Selvam[124], in particular has shown that round - off error approximately doubles for each iteration of finite precision iterative computations and enters the mainstream computation within 50 iterations and thereafter the computed solution gives only the round - off error growth structure. Numerical weather/climate prediction models incorporate thousands of iterative computations in numerical integration schemes and therefore the model solutions will only represent the dynamical evolution of round - off error growth.

The accurate modelling of weather/climate phenomena therefore requires alternative concepts and computational techniques. Theoretical concepts and analytical techniques developed so far in the multidisciplinary new science of nonlinear dynamics and chaos have to be adapted and incorporated in classical meteorological theory for realistic prediction of weather phenomena.

## Acknowledgement

The author is grateful to Dr. A. S. R. Murty for encouragement during the course of this study.